\documentclass[%
 reprint,
showpacs,
 amsmath,amssymb,
 aps,
]{revtex4-1}

\usepackage{graphicx}
\usepackage{dcolumn}
\usepackage{bm}

\usepackage[
margin=0.65in,top=0.1in,
]{geometry}

\begin{document}

\preprint{APS/123-QED}

\title{Exciton-Phonon Coupling and Band-Gap Renormalization in Monolayer WSe$_2$}
\author{Himani Mishra${}^\ddagger$}
\author{Anindya Bose${}^\ddagger$}%
\author{Amit Dhar$^{\dagger}$}
 \author{Sitangshu Bhattacharya${}^\ddagger$}
 \email{Corresponding Author's Email: sitangshu@iiita.ac.in}
\affiliation{${}^\ddagger$Nanoscale Electro-Thermal Laboratory, Department of Electronics and Communication Engineering, \\ $\&$ \\
${}^\dagger$Department of Information Technology, \\
Indian Institute of Information Technology-Allahabad, Uttar Pradesh 211015, India}
\begin{abstract}
Using a fully ab-initio methodology, we demonstrate how the lattice vibrations couple with neutral excitons in monolayer WSe$_{2}$ and contribute to the non-radiative excitonic lifetime. We show that only by treating the electron-electron and electron-phonon interactions at the same time it is possible to obtain an unprecedented agreement of the zero and finite-temperature optical gaps and absorption spectra with the experimental results. The bare energies were calculated by solving the Kohn-Sham equations, whereas G$_{0}$W$_{0}$ many body perturbation theory was used to extract the excited state energies. A coupled electron-hole Bethe-Salpeter equation was solved incorporating the polaronic energies to show that it is the in-plane torsional acoustic phonon branch that contributes mostly to the A and B exciton build-up. We find that the three A, B and C excitonic peaks exhibit different behaviour with temperature, displaying different non-radiative linewidths. There is no considerable transition in the strength of the excitons with temperature but A-exciton exhibits darker nature in comparison to C-exciton. Further, all the excitonic peaks redshifts as temperature rises. Renormalization of the bare electronic energies by phonon interactions and the anharmonic lattice thermal expansion causes a decreasing band-gap with increasing temperature. The zero point energy renormalization (31 meV) is found to be entirely due to the polaronic interaction with negligible contribution from lattice anharmonicites. These findings may find a profound impact on electronic and optoelectronic device technologies based on these monolayers.
\end{abstract}
\maketitle

\section{\label{sec:level1}Introduction}
Transitional metal dichalcogenide (TMDC) family and their monolayer (ML) counterparts have distinguished themselves as a perfect platform to understand various finite temperature quantum many body phenomena \cite{Kogar2017, Srivastava2015, Moratalla2016, Hao2016}. Among them, the optical absorption spectra characterize the fundamental band-edges and carrier lifetime. When a photon with energy larger than this gap is incident, electron changes it's state from valance band to conduction band, thereby leaving a hole in the former. If the external screening between them is weak, then this particle-hole pairs for a sufficiently long time via Coulombic attraction and mimics a two-dimensional hydrogen-like atom with quantized energy states. Fingerprints of such quasi-particles (QPs) can be obtained from peaks in the optical absorption spectra determined from non-linear two-photon photoluminescence (2PPL) and angle-resolved photoemission spectroscopy (ARPES) \cite{Suzuki2014, Alidoust2014, Riley2014, He2014, Ye2014, Wang2015} measurements. The energies corresponding to these peaks are of central importance as they determine the conditions of maximum generation and recombination. However, this QP or exciton dance is strongly affected by the presence of lattice vibrations which make them dissociate ultimately. Determination of this lifetime therefore stands out as a bottle-neck to understand the exciton-phonon dynamics in optical devices based on these materials.\\
Intrinsic exfoliated ML WSe$_{2}$, when illuminated by such photons display sharp and strongest bright neutral excitonic ground state peaks A and B at 1.65 eV and 2.05 eV respectively with a giant binding energy of 0.37 eV at 300 K \cite{He2014, Ahna2017}. These along with other excited excitonic states can be well captured from 2PPL measurements, done along $\Gamma$-$\textbf{K}$ and $\Gamma$-$\textbf{K}^{\prime}$ of the Brillouin route, confirming that ML WSe$_{2}$ possess a direct optical band gap. Conclusive evidence about the giant valance spin-orbit splitting (SoS) of 513$\pm$10 meV at $\textbf{K}$ and $\textbf{K}^{\prime}$ arising due to the lack of crystal inversion symmetry can also be obtained from the ARPES measurements \cite{Le2015}. As it is difficult to capture the the conduction band or its splitting profile directly from ARPES, one necessarily probes for the presence of a dark exciton. In these MLs, a dark exciton is formed when an electron pairs to a hole with anti-parallel spin (intra-valley) and parallel spin (inter-valley). Fortunately, due to the presence of time-reversal (TR) symmetry $\left(E\left(\textbf{K}\uparrow\right)=E\left(\textbf{K}^{\prime}\downarrow\right)\right)$, the spin and momenta at these two valleys are coupled \cite{Srivastava2015, Mak2012, Jones2013}. This means that at $\textbf{K}$ the top two valance bands separately possess spin-up and down, while the two lowest conduction bands separately possess spin-down and up respectively. A complementary scenario is then obtained at $\textbf{K}^{\prime}$ by exploiting the TR symmetry. Therefore, a dark exciton can only form when a hole at top valance band at $\textbf{K}$ pairs with an electron at the lowest conduction band of $\textbf{K}$ (intra-valley) and $\textbf{K}^{\prime}$ (inter-valley) \cite{Poem2010, Malic2018}. Conversely, since a bright exciton is formed only when an electron pairs to a hole with a parallel spin (intra-valley) and an anti-parallel spin (inter-valley), the difference in the bright-dark excitonic energies provides the splitted energy difference between the conduction bands. This is found to be 30$\pm$5 meV for ML WSe$_{2}$ \cite{Zhang2015}. In MoX$_{2}$ (X=Se, Te, S) MLs, this energy is found to be negative suggesting that the dark excitons house themselves in higher energies, beyond the first ground state bright exciton.\\ 
Many body perturbation theory (MBPT) has been quite successful in explaining these excitonic dynamics and has shown convincing results done on Si and other hexagonal family members \cite{Marini2008,Alejandro2013,Shu2016, Galvani2016,Franti2013,Antonius2018}. However a quantitative understanding about the exciton-phonon coupling in ML WSe$_{2}$ still needs to be addressed. Particularly, it is the excitonic non-radiative widths which are most difficult to capture through experiments and therefore we use the MBPT as a tool to calculate it. Here we demonstrate the integrated effect of the electron-electron and electron-phonon interactions on the electronic and optical properties of intrinsic ML WSe$_{2}$. We show that under such scenario, there is an unprecedented agreement of the zero and finite-temperature optical gaps and absorption spectra with the experimental results. We start with the evaluation of bare electronic energies by first solving the Kohn-Sham set of equations using the density functional theory (DFT). MBPT was then used to compute the QP energies on the top of this by switching on the dynamic electron-electron correlation. This eventually fixed the energy gap which was otherwise grossly underestimated using only DFT. In order to get the renormalized excitonic energies, absorption spectra and lifetimes, we solve a coupled electron-hole Bethe-Salpeter equation (BSE) that included the polaronic energies extracted from the density functional perturbation theory (DFPT) calculations giving an excellent description of the phonon frequencies and coupling strengths. The methodology adopted here is therefore a fully $ab$-$initio$ and demands no external parameters.
\section{Theory}
In intrinsic crystals, there are three types of coupling that mostly controls the temperature dependent absorption spectra. The first one is the coupling between electronic and atomic degrees of freedom. MBPT provides useful information about the zero point renormalization (ZPR) of the widths not present in the bare-particle states $\left|n,\textbf{k}\right\rangle$. The electron-phonon (EP) matrix elements \cite{Fan1950} 
\begin{eqnarray}
g_{n^{\prime}n\textbf{k}}^{\textbf{q}\lambda}=\sum_{\alpha s}\left\langle n,\textbf{k}\left|\nabla_{\alpha s}\phi_{scf}\right|n^{\prime},\textbf{k}+\textbf{q}\right\rangle \nonumber \\
\times \sum_{\textbf{q}\lambda}\left(\frac{1}{2M_{s}\omega_{\textbf{q}\lambda}}\right)^{\frac{1}{2}}e^{-iq\cdot\tau_{s}}\epsilon^{\ast}\left(\frac{\textbf{q}\lambda}{s}\right)
\end{eqnarray}
describes the electron scattering probability amplitudes from \textbf{k} to \textbf{k}+\textbf{q} due to the emission or absorption of a phonon with transferred momenta $\textbf{q}$ in branch $\lambda$. Here, the self-consistent potential $\phi_{scf}$ is first obtained by calculating the charge density from DFT. DFPT is then used to calculate the first-order derivative of $\phi_{scf}$ with respect to atomic displacements $\alpha$ and consequently the dynamical matrices. For this, the entire Brillouin zone (BZ) is sampled by a number of random \textbf{q}-points. $\tau_{s}$ is the location of mass $M$ of the $s^{th}$ atomic species in the unit cell and $\epsilon^{*}\left(\frac{\textbf{q}\lambda}{s}\right)$ are the polarization vectors. The states  $\left|\textbf{k}+\textbf{q}\right\rangle$ are finally obtained via a non-self-consistent calculation done on the same regular grid. Using the Matsubara Green's function, one can write the polaronic self-energies after an analytic continuation on the real-axis as \cite{Fan1950}
\begin{eqnarray}
\sideset{}{_{n\textbf{k}}^{Fan}}\sum\left(\omega\right)&=&\sum_{n^{\prime}\textbf{q}\lambda}\left|g_{n^{\prime}n\textbf{k}}^{\textbf{q}\lambda}\right|^{2}\left[\frac{N\left(\omega_{\textbf{q}\lambda}\right)+1-f_{n^{\prime}\textbf{k}-\textbf{q}}}{\omega-\epsilon_{n^{\prime}\textbf{k}-\textbf{q}}-\omega_{\textbf{q}\lambda}-i0^{+}} \right. \nonumber \\ &+& \left. \frac{N\left(\omega_{\textbf{q}\lambda}\right)+f_{n^{\prime}\textbf{k}-\textbf{q}}}{\omega-\epsilon_{n^{\prime}\textbf{k}-\textbf{q}}+\omega_{\textbf{q}\lambda}-i0^{+}}\right]
\end{eqnarray}
Here $N$ and $f$ are the phonon and electron distribution functions respectively and 0$^{+}$ is used to make the contour integral vanish over the half circle of the upper half plane. The second order EP matrix elements stems from the renormalized screening due to the atomic motion and are calculated by invoking the translational invariance symmetry. The frequency independent self-energies can then be written as \cite{Cannuccia2011}
\begin{equation}
\sideset{}{_{n\textbf{k}}^{DW}}\sum=-\sum_{\textbf{q}\lambda}\sum_{n^{\prime}}\frac{1}{2}\frac{\left|\Lambda_{nn^{\prime}\textbf{k}}^{\textbf{q}\lambda}\right|^{2}}{\epsilon_{n\textbf{k}}-\epsilon_{n^{\prime}\textbf{k}}}\left[2N\left(\omega_{\textbf{q}\lambda}\right)+1\right]
\end{equation}
$\Lambda_{nn^{\prime}\textbf{k}}^{\textbf{q}\lambda}$ are the corresponding couplings.\\
Equations (2)-(3) are the Fan and Debye-Waller (DW) self-energies used to compute the electronic Green's function $G_{n\textbf{k}}^{EP}\left(\omega\right)=\left[\omega-\epsilon_{n\textbf{k}}-\sideset{}{_{n\textbf{k}}^{Fan}}\sum\left(\omega\right)-\sideset{}{_{n\textbf{k}}^{DW}}\sum\left(\omega\right)\right]^{-1}$. The real part of this pole is the renormalized QP energies while the imaginary part corresponds to the polaronic lifetime. Assuming that the bare energies $\epsilon_{n\textbf{k}}$ are far from the poles of the real or imaginary Fan self-energies, one does a Taylor's expansion about $\epsilon_{n\textbf{k}}$ and a renormalization weight factor $Z_{n\textbf{k}}^{EP}$ $\left(0<Z_{n\textbf{k}}^{EP}<1\right)$ is therefore assigned to each $\left|n,\textbf{k}\right\rangle$ state resulting in
\begin{equation}
\Delta E_{n\textbf{k}}^{EP}-\epsilon_{n\textbf{k}}\approx Z_{n\textbf{k}}^{EP}\Re\left[\sideset{}{_{n\textbf{k}}^{Fan}}\sum\left(\omega\right)+\sideset{}{_{n\textbf{k}}^{DW}}\sum\left(\omega\right)\right]
\end{equation}
with $Z_{n\textbf{k}}^{EP}=\left[1-\left.\frac{\partial}{\partial\omega}\Re\sum_{n\textbf{k}}^{Fan}\left(\omega\right)\right|_{\omega=\epsilon_{n\textbf{k}}}\right]^{-1}$. Equation (4) is the dynamical HAC (Heine, Allen and Cardona) \cite{Cardona2005} theory and represents a finite zero-point energy even when $T\rightarrow0$ K.  This way the uncertainty principle is is also satisfied with the added advantage that the exciton absorption spectra does not demand any fitting-broadening parameter. When $\epsilon_{n\textbf{k}}$ are very far from the poles of real and imaginary Fan self-energies such that $\frac{\partial}{\partial\omega}\Re\sum_{n\textbf{k}}^{Fan}\left(\omega\right)=\frac{\partial}{\partial\omega}\Im\sum_{n\textbf{k}}^{Fan}\left(\omega\right)=0$, only virtual electronic scatterings are allowed and the Taylor's expansion is computed upto the zeroth order. This is known as static limit or the on-the-mass-shell approach ($Z_{n\textbf{k}}\rightarrow$1). The imaginary part of $G_{n\textbf{k}}^{EP}\left(\omega\right)$ also reflects the spectral function (SF) $\mathsf{A}_{n\textbf{k}}^{EP}\left(\omega\right)=\frac{1}{\pi}\left|\Im G_{n\textbf{k}}^{EP}\left(\omega\right)\right|$. A weak coupling results in a sharp SF centred at $\Delta E_{n\textbf{k}}^{EP}$. As the coupling gets stronger the Lorentzian SF peak start becoming more asymmetric and dwarf. Physically, $Z_{n\textbf{k}}^{EP}$ signifies the fraction of the positive charge that a bare electron takes away from the QP cloud and form satellite-peaks. Comparing the energy difference between the main and the satellite peak with the Debye energy, the validity of QP assumption can be verified. Equation (4) can also be used to understand which phonon mode actually contributed to the non-radiative excitonic linewidth. This can be calculated from the Eliashberg function for each state $\left|n,\textbf{k}\right\rangle$ \cite{Mahan2014}
\begin{equation}
g_{n\textbf{k}}^{2}F\left(\omega\right)=\sum_{q\lambda}\frac{\partial E_{n\textbf{k}}^{EP}}{\partial N\left(\omega_{q\lambda}\right)}\delta\left(\omega-\omega_{q\lambda}\right)
\end{equation}\\
and mapping it to the corresponding phonon dispersion.\\
In addition to EP coupling, lattice thermal expansion (LTE) also modifies the bare-particle energies and can be of same order compared to the former \cite{Villegas2016}. The energy renormalization within a variable volume quasi-harmonic approximation (QHA) can be obtained by minimizing the Helmholtz free energy \cite{Mounet2005}
\begin{eqnarray}
F\left(\left\{ a_{j}\right\} ,T\right)-\epsilon\left(\left\{ a_{j}\right\} \right)=\sum_{q,\lambda}\frac{\hbar\omega_{q,\lambda}\left(\left\{ a_{j}\right\} \right)}{2} \nonumber \\
+k_{B}T\sum_{q,\lambda}\mathrm{ln}\left[1-\mathrm{exp}\left(-\frac{\hbar\omega_{q,\lambda}\left(\left\{ a_{j}\right\} \right)}{k_{B}T}\right)\right]
\end{eqnarray}
The first-term on right is the vibrational zero-point energy while the second term in the left is the DFT-bare energy performed with respect to the geometrical degrees of freedom {$a_{j}$} and the sums extending the BZ. Equation (4) when combined to Eq. (6) gives the net ZPR due to LTE and EP coupling.\\
The second coupling is the dynamic long range electron-electron (EE) correlation and is responsible for opening of the QP electronic energy gap or simply the G$_{0}$W$_{0}$ gap. Within the linear response theory, the dynamic self-energy can be expressed from the density-density response function as \cite{Mahan2014}
\begin{widetext}
\begin{equation}
\sum_{n\textbf{k}}\left(\omega\right)=i\sum_{m}\int_{BZ}\frac{d\textbf{q}}{\left(2\pi\right)^{3}}\sum_{\textbf{GG}^{\prime}}\frac{4\pi}{\left|\textbf{q}+\textbf{G}\right|^{2}}\rho_{nm}\left(\textbf{k},\textbf{q},\textbf{G}\right)\rho_{nm}^{\ast}\left(\textbf{k},\textbf{q},\textbf{G}\right)\int d\omega^{\prime}\textbf{G}_{m\textbf{k}-\textbf{q}}^{0}\left(\omega-\omega^{\prime}\right)\varepsilon_{\textbf{GG}^{\prime}}^{-1}\left(\textbf{q},\omega^{\prime}\right)
\end{equation}
\end{widetext}
in which $m$ spans all the occupied to unoccupied bands and $\textbf{G}$ are the G-vectors of the plane-waves in the Fourier transformed plane. $\varepsilon_{\textbf{GG}^{\prime}}^{-1}$ is the microscopic frequency dependent dielectric function which can be efficiently calculated using the Godby-Needs plasmon-pole approximation model   \cite{Godby1989} to solve the dynamic screening $W$ at the random-phase approximation (RPA) level. By including the static exchange part of the self-energy, the QP energies can be obtained as $\Delta E_{n\textbf{k}}^{EE}=\epsilon_{n\textbf{k}}+Z_{n\textbf{k}}^{EE}\left\langle \psi_{n\textbf{k}}\left|\sum\left(\epsilon_{n\textbf{k}}\right)-V_{xc}\right|\psi_{n\textbf{k}}\right\rangle $ in which $V_{xc}$ is obtained by solving the Kohn-Sham equations. It is then clear that the corrections to the band gap due to the EE correlation is orders of magnitude higher than the EP interactions. Similar to EP case, the SF here is again the imaginary Green's propagator that can be captured well by ARPES measurements. \\
The third is the dynamic long range electron-hole ($e$-$h$) coupling describing the two particle Green's propagator. The corresponding equation of motion is the BSE \cite{Fetter2013}. Under the frozen-atom (FA) approximation (i.e., no atomic vibrations), the BS Hamiltonian is a Hermitian matrix in the $e$-$h$ pair basis with eigenstates $\left|\varphi_{FA}\left(T\right)\right\rangle $. The excitonic eigenenergies can then be obtained by formulating the corresponding Hamiltonian in the from  \cite{Marini2008}
\begin{equation}
H_{ee^{\prime},hh^{\prime}}^{FA}=\left(E_{e}-E_{h}\right)\delta_{eh,e^{\prime}h^{\prime}}+\left(f_{e}-f_{h}\right)K_{ee^{\prime},hh^{\prime}}
\end{equation}
in which $K_{ee^{\prime},hh^{\prime}}$ is BS kernel and is the sum of a repulsive (positive) bare Coulomb exchange term and an  attractive (negative) direct $e$-$h$ screened interaction term. The former stems from to the variation of Hartree potential and is responsible for spin-singlet/triplets splitting while the later is a long-range and responsible for formation of excitonic bound states. $(E/f)_{e/h}$ are the corresponding quasi-$e$/$h$ energies and Fermi occupation numbers. When lattice vibrations and LTE effects are present, $E_{e}$ and $E_{h}$ broadens to $\Delta E_{e}$ and $\Delta E_{h}$ respectively and makes the BS Hamiltonian a non-Hermitian operator. However, in most semiconductors the QP corrections owing to LTE are much less compared to $\Delta E_{nk}^{EP}$, except in some special cases resulting in an anomalous band gap dependency on temperature \cite{Villegas2016}. This is yet not reported in monolayer TMDCs. Relaxing therefore only LTE contributions, the excitonic energy eigenvalues can be written in the form \cite{Marini2008}\\
\begin{eqnarray}
E_{\varphi}\left(T\right)=\left\langle \varphi\left(T\right)\left|H^{FA}\right|\varphi\left(T\right)\right\rangle +\sum_{e,h}\left|A_{e,h}^{\varphi}\left(T\right)\right|^{2} \nonumber \\ 
\times\left[\Delta E_{e}\left(T\right)-\Delta E_{h}\left(T\right)\right]
\end{eqnarray}\\
It is then straight forward to extract the real and imaginary parts from Eq. (9) as
\begin{eqnarray}
\Re\left[\Delta E_{\varphi}\left(T\right)\right]=&&\left\langle \varphi\left(T\right)\left|H^{FA}\right|\varphi\left(T\right)\right\rangle -\left\langle \varphi_{FA}\left|H^{FA}\right|\varphi_{FA}\right\rangle\nonumber\\
&& +
\int d\omega\Re\left[g^{2}F_{\varphi}\left(T\right)\right]\left[N\left(\omega,T\right)+\frac{1}{2}\right]
\end{eqnarray}
and 
\begin{equation}
\Im\left[E_{\varphi}\left(T\right)\right]=\int d\omega\Im\left[g^{2}F_{\varphi}\left(T\right)\right]\left[N\left(\omega,T\right)+\frac{1}{2}\right]
\end{equation}
The difference $g^{2}F_{\varphi}\left(T\right)=\sum_{e,h}\left|A_{eh}^{\varphi}\left(T\right)\right|^{2}\left[g^{2}F_{e}\left(\omega\right)-g^{2}F_{h}\left(\omega\right)\right]$ represent the exciton-phonon coupling function and $\Delta E_{\varphi}\left(T\right)=E_{\varphi}\left(T\right)-E_{\varphi}^{FA}$. 
The temperature dependent macroscopic dielectric function in the long wavelength limit is therefore 
\begin{equation}
\varepsilon_{M}\left(\omega,T\right)=-\frac{8\pi}{\Omega}\sum_{\varphi}\left|O_{\varphi}\left(T\right)\right|^{2}\Im\left(\omega-E_{\varphi}\left(T\right)\right)^{-1}
\end{equation}
$\Im\varepsilon_{M}\left(\omega,T\right)$ defines the complete absorption spectra with the exciton oscillator strength $O_{\varphi}\left(T\right)=\left\langle n\textbf{k}\left|exp\left(i\boldsymbol\kappa\cdot\mathbf{r}\right)\left(\left[\left.\left|\varphi\left(T\right)\right.\right\rangle -\left.\left|\varphi^{FA}\left(T\right)\right.\right\rangle \right]\right)\right.\right.$ in which $\boldsymbol\kappa$ is the electric polarization vector direction and $\Omega$ as the volume. A more elaborate theoretical treatment on Hedin's GW formalism and BSE can be found elsewhere \cite{Hedin1965, Fetter2013, Rohlfing2000}.
\section{Results and Discussions}
WSe$_{2}$ belong to the space group P63/mmc. In order to replicate a full all-electron potential within the atoms, a norm-conserving and fully relativistic pseudopotential was first generated \cite{Hamann2013}. The 5s, 5p and 4f semi-core orbitals were included for W along with 5d and 6s valence electrons. PBE exchange-functional was selected because of two reasons: (a) after testing with variety of other functionals the convergence of lattice constant here is found to be in good agreement compared to the experimental data (with a slight overestimation of 0.04 $\mathring{\mathrm{A}}$) and (b) exchange-functional like LDA tends to underestimate the electron-phonon interactions by 30$\%$ \cite{Antonius2015}. A unit cell of WSe$_{2}$ constituting 3 atoms was therefore first generated. The structure was truncated in the out-of-plane direction using a vacuum-slab-vacuum profile of 25 $\mathring{\mathrm{A}}$ in each side to avoid the Coulombic interaction between repeated images. $\mathtt{PWscf}$ code \cite{Giannozzi2009} was then used to solve the Kohn-Sham equations. After testing the variety of kinetic cut-off energies, convergence was found to be achieved at 90 Ry for both the atomic species (See Supplementary Information). A $\Gamma$ centred Monkhorst-Pack scheme is then implemented for sampling of BZ with a dense-grid of 18$\times$18$\times$1. The structure is then allowed to relax within the variable cell configuration. The cell sizes as well as atomic coordinates were continuously updated along the minimum slope direction specified by the locally optimized coordinates calculated from the Hellmann-Feynman theorem. This minimizes the total energy of the unit cell with the forces less than 10$^{-4}$ eV$\mathring{\mathrm{A}}^{-1}$. Various sets of fully-relativistic, same semi core-corrected orbitals and norm-conserving pseudopotentials were generated to realize the experimentally evaluated in-plane lattice constant and is shown below in Table 1.
\begin{table}
  \caption{In-plane lattice parameter and band-gap with various exchange functionals}
  \label{tbl:notes}
  \begin{tabular}{lll}
    \hline
     Exchange functionals   & a ($\mathring{\mathrm{A}}$)   & E$_{g}$ (eV) \\
    \hline
    PZ    & 3.2468	& 1.387\\
    PBE   & 3.3175	& 1.26\\
    REVPBE   & 3.3365	& 1.217\\
    BP   & 3.3333	& 1.2269\\
    \hline
  \end{tabular}

  \textsuperscript{}Experimental lattice constant: 3.28 ($\mathring{\mathrm{A}}$) \cite{Wilson1969}
\end{table}
Among all others, we found PBE exchange functional to correctly describe the relaxed in-plane lattice constant. A further optimization was done without varying the cell volume but within the same force limits. A two-spinor wave function was finally expanded in the plane wave basis set along with a non-collinear and SoC calculation was deployed to carry out a self-consistently evaluated charged densities. A non self-consistently evaluated states calculation along the regular grid was finally done to interpolate the bare energy bands. \\
In Figs. (1 a-b), $\mathsf{V_{1}}$ and $\mathsf{V_{2}}$ at $\mathsf{K}$ are the splitted valance bands (Zeeman-like) due to the SoC in which both of them are mainly populated with 5d orbital electrons from W-atom. The spin occupancy due to the L-S coupling results in an orbital angular momentum ($l=2$) and spin quantum number $m_{s}=+\frac{1}{2}$ for $\mathsf{V_{1}}$ and $-\frac{1}{2}$ for $\mathsf{V_{2}}$ respectively. The giant spin-orbit splitting between these two is found to be $\mathsf{\left|\mathrm{\Delta E_{v}^{SoC}}\right|}=466$ meV, while the energy difference of $\mathsf{V_{1}}$ between $\mathsf{K}$-$\mathsf{\Gamma}$ direction is -492 meV. Further, the SoC results in a splitting of conduction band $\mathsf{C_{1}}$ and $\mathsf{C_{2}}$ at $\mathsf{K}$ with $\mathsf{\left|\mathrm{\Delta E_{C}^{SoC}}\right|}=37$ meV. 
\begin{figure*}
\includegraphics[width=1.5\columnwidth]{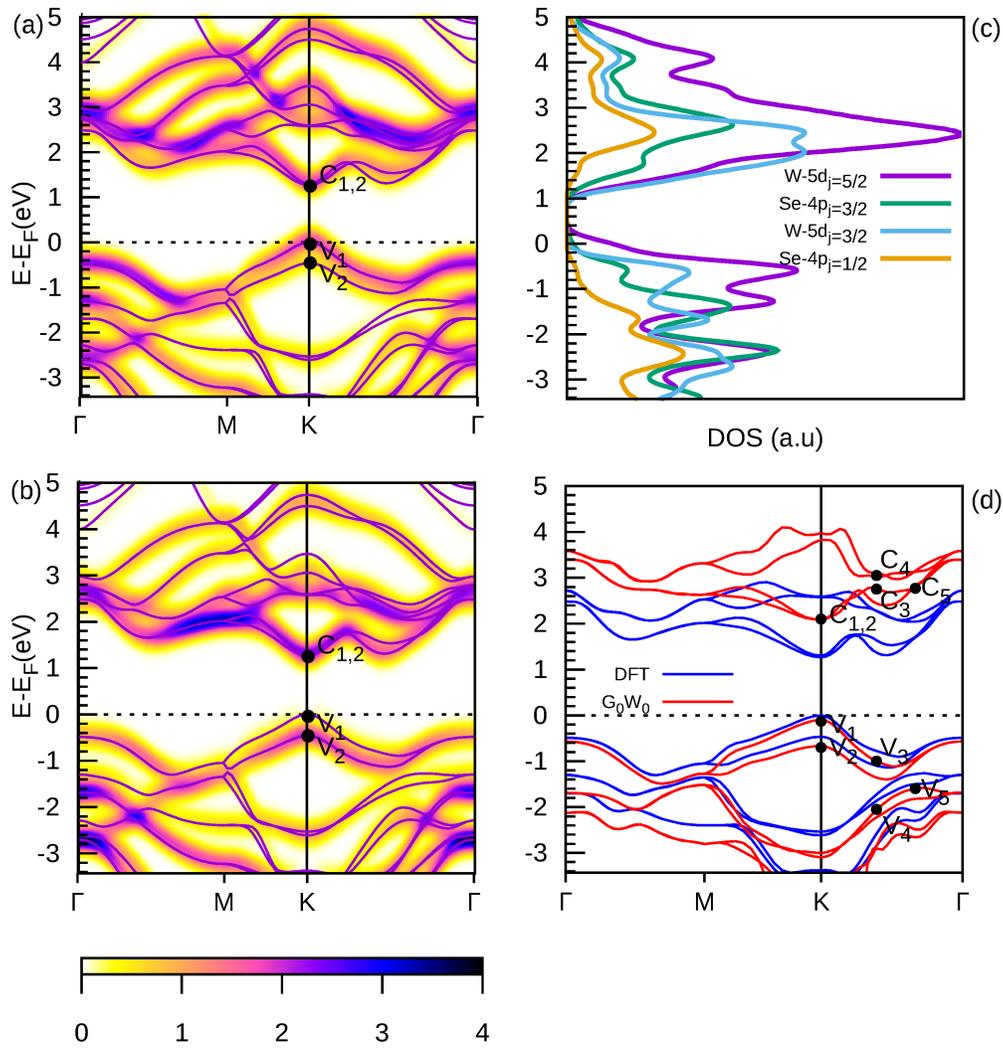}
\caption{(a-b) Bare electronic energy dispersions in ML WSe$_{2}$ along the BZ. Spin is resolved and projected onto each state with ($l=$2) and exhibiting $m_{s} = +\frac{1}{2}$ occupancy in (a) and $-\frac{1}{2}$ occupancy in (b). These occupancies are shown by more darker areas. The partial density-of-states is shown in (c) for 5d and 4p orbitals of W and Se respectively mostly influencing the total angular momentums. The anti-parallel spins is clearly visible near the bottom of conduction and top of the valance bands. Excited state G$_{0}$W$_{0}$ dispersions are shown in (d) with a direct band gap of 2.19 eV. $\mathsf{C}$s and $\mathsf{V}$s represents electronic transitions during optical excitations.}
\end{figure*}
The lowest valley conduction band in the $\mathsf{K}$-$\mathsf{\Gamma}$ direction is almost degenerate with $\mathsf{C_{1}}$ at $\mathsf{K}$ and is located above by 44 meV (See Supplementary Information for a magnified diagram). These band-edges therefore results in a direct bare energy gap (E$_{g}$) of almost =1.26 eV. The partial-density of states contributed by different atomic orbitals and spins is shown in Fig. (1-c) in which the conduction bands $\mathsf{C_{1}}$ and $\mathsf{C_{2}}$ are mainly found to build up by $5d_{j=\frac{3}{2}}$ (spin-down) and $5d_{j=\frac{5}{2}}$ (spin-up) of W atom respectively. This is complementary to the valance band spin distribution. This anti-parallel spin occupancy in both splitted valance and conduction bands is due to the broken inversion symmetry. Interestingly, in case of ML MoX$_{2}$ (X=S, Se, Te) a parallel spin polarizations in $\mathsf{V_{1}}$, $\mathsf{C_{1}}$ and $\mathsf{V_{2}}$, $\mathsf{C_{2}}$ are found \cite{Suzuki2014}.\\
\begin{figure*}[!ht]
\includegraphics[width=2.0\columnwidth]{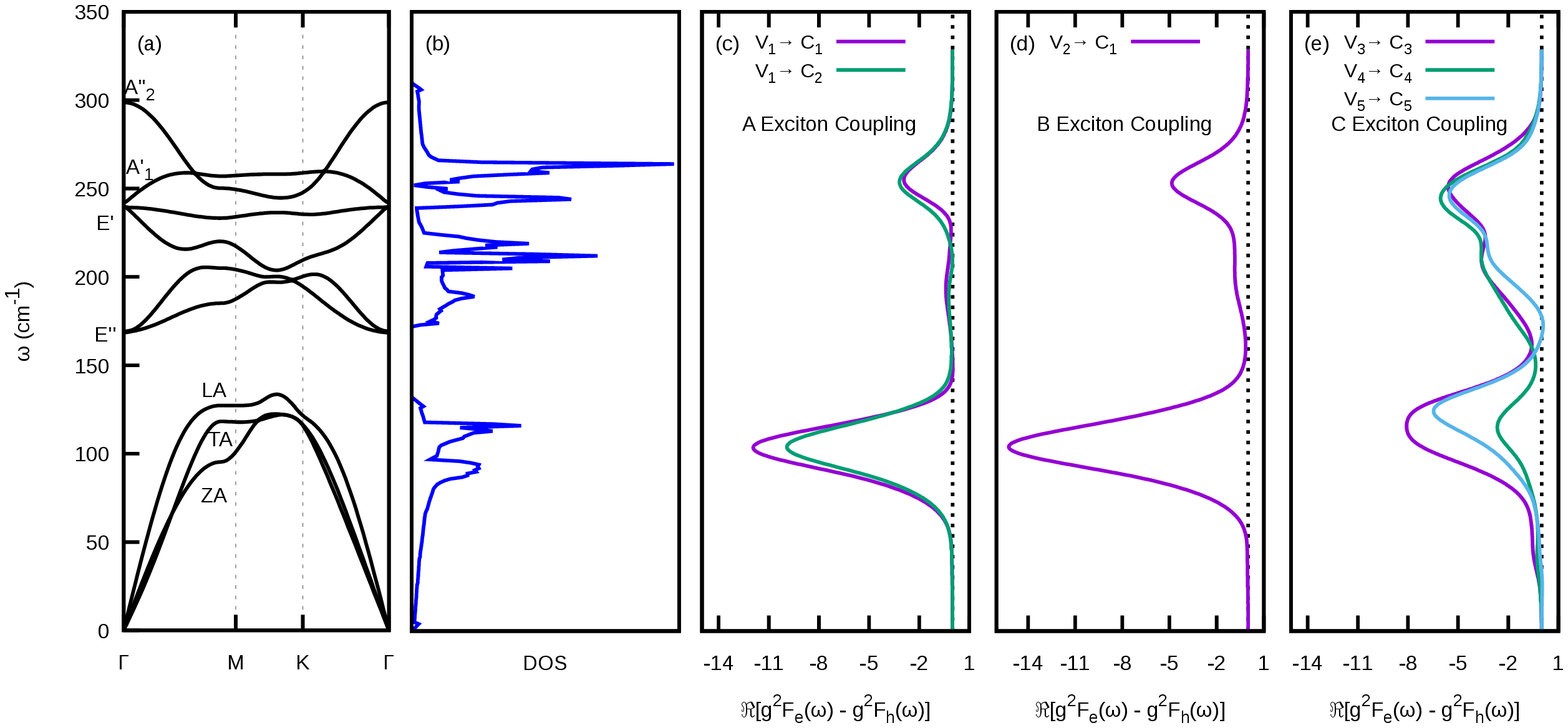}
\caption{Phonon (a) dispersion and (b) density-of-states in ML WSe$_{2}$ along the BZ. The real part of the difference between the conduction and valance band Eliashberg functions along the lattice frequency of vibrations calculated for (c) $\left(\mathsf{V}_{1},\mathsf{C}_{1}\right)$, $\left(\mathsf{V}_{1},\mathsf{C}_{2}\right)$ (responsible for A exciton build up), (d) $\left(\mathsf{V}_{2},\mathsf{C}_{1}\right)$ (responsible for B exciton build up) and (e) $\left(\mathsf{V}_{3},\mathsf{C}_{3}\right)$, $\left(\mathsf{V}_{4},\mathsf{C}_{4}\right)$ and $\left(\mathsf{V}_{5},\mathsf{C}_{5}\right)$ (responsible for C-exciton build up). All the differences are negative suggesting a shrinkage of band gap as temperature rises.}
\end{figure*}
To capture the excitation results, the excited state corrections G$_{0}$W$_{0}$ method was employed. The excited state computation was carried out by using the YAMBO code \cite{Andrea2009} and involved a careful calculation of the QP energies with a dynamic dielectric screening function evaluated with the general plasmon-pole model. One of the two plasmon-pole frequency was chosen to be 27.2 eV beyond which dielectric function remained unchanged. The exchange part in the static self-energy (Hartree-Fock) was summed over 20065 $\textbf{G}$-vectors. 150 bands with the number of unoccupied bands 1.5 times of the number of occupied bands were used for summation of the polarization function (Eq. (7)). The QP corrections was done in the lowest 5 empty conduction bands and highest 6 valence bands along the BZ and was found sufficient to obtain a converged direct band gap of 2.19 eV in the G$_{0}$W$_{0}$ calculation. The energy cut-off for the response function was set to 5 Ry. An important ingredient was to manage the divergence of the Coulomb potential at small $\textbf{q}$ appearing in the exchange self-energy (Eq. (7)), BSE, etc. for reduced geometries. To fix this, a random integration method \cite{Olivia1998} was applied which assumes a smooth momenta integrand function (only for oscillators and occupation numbers) about $\textbf{q}$ in each region of the BZ without changing the potential itself. 3000000 random points were incorporated in the calculation in order to evaluate the Coulomb integrals upto 100 $\textbf{G}$-vectors of the Coulomb potential in small BZ regions. A Monte Carlo technique was then used to evaluate this integral defined within a box-structure extending 25 $\mathring{\mathrm{A}}$ on either side of the ML WSe$_{2}$. This truncated the Coulombic potential between the repeated images and a faster convergence was achieved. The inclusion of the dynamic EE correlation ((Eq. (7)) then opens up E$_{g}$ to 2.19 eV which is in excellent agreement to the reported range of 2.02 - 2.35$\pm$0.2 eV for exfoliated WSe$_{2}$ MLs \cite{He2014,Wang2015}. Comparing both the G$_{0}$W$_{0}$ and DFT results, it is evident that the difference $Z_{n\textbf{k}}^{EE}\left\langle \psi_{n\textbf{k}}\left|\sum\left(\epsilon_{n\textbf{k}}\right)-V_{xc}\right|\psi_{n\textbf{k}}\right\rangle $ essentially provides a rigid energy shift only (with $\Re Z^{EE}$=77$\%$ and 76$\%$ for $\mathsf{V_{1}}$ and $\mathsf{C_{1}}$ respectively), concluding that this dynamic EE correlated QP corrections are rather weakly dependent on $\textbf{k}$-vector directions. It is also evident that the change in the energies of the unoccupied states after G$_{0}$W$_{0}$ corrections are considerably larger when compared to the occupied ones and are mainly due to the change in the binding energies which results in an increase in the valence bandwidth. In the case of unoccupied states, the low lying conduction bands formed now have larger QP corrections as they are highly localized in nature; supporting the fact that self-interaction errors in DFT are severe for localized states therefore producing infinite lifetime. Further, the two low lying non-degenerate conduction bands at $\mathsf{K}$ after G$_{0}$W$_{0}$ corrections acquires 40 meV separation. This is in excellent agreement with the experimentally evaluated splitting of the conduction bands (30$\pm$5 eV) found by probing the dark excitonic states \cite{Zhang2015}. The almost degenerate satellite bands obtained in Kohn-Sham calculations are now lifted up to 393.8 meV while the valance band splitting opens up to $\mathsf{\left|\mathrm{\Delta E_{v}^{SoC,\mathrm{G}_{0}\mathrm{W}_{0}}}\right|}=572$ meV.\\
DFPT was now used to understand the lattice vibrations and EP self-energies. Bulk WSe$_{2}$ belongs to $D_{4h}\left(\overline{6}2m\right)$ symmorphic group. Thus, a 12$\times$12$\times$1 $\textbf{k}$ sampling of the BZ was used for ML WSe$_{2}$ and was found sufficient for the evaluation of the Fan and Debye-Waller self-energies with a rigid self-consistent error threshold below 10$^{-16}$ Ry and a single iteration mixing-factor of 0.8 Ry that continuously updated $\phi_{scf}$. Additionally, a phonon broadening of 1 meV is used in the analysis of Eliashberg function to capture the phonon mode contributions. In the ML counterpart, the group symmetry reduces to $D_{3h}\left(\overline{6}2m\right)$ where the nine modes at $\Gamma$ decomposes into four irreducible representations $A_{2}^{\prime\prime}$, $E^{\prime}$, $E^{\prime\prime}$ and $A_{1}^{\prime}$. The dispersion and the density-of-states have been exhibited in Figs. (2 a-b). Modes $E^{\prime}$, $E^{\prime\prime}$ and $A_{1}^{\prime}$ are equivalent to $E_{2g}$, $E_{1g}$ and $A_{1g}$ in the bulk case respectively. Both the in-plane longitudinal acoustic (LA) mode and out-of plane optical (ZO) mode $A_{2}^{\prime\prime}$ are IR (infra-red) active. The ZA and TA motions $E^{\prime}$ are both degenerate, IR and Raman active. The two mid frequency modes $E^{\prime\prime}$ are in-plane, degenerate and only Raman active. Further, the two modes $E^{\prime}$ in the lower optical regions are degenerate with both IR and Raman active, while the out-of phase ZO modes $A_{1}^{\prime}$ (Se-Se) and $A_{2}^{\prime\prime}$ (Se-W-Se) are IR and Raman active respectively. The first- and second-order E-P matrix elements (Eqs. (2)-(3)) were then calculated by sampling the entire BZ into 113 random transferred-momenta grid and 288 random $\textbf{k}$-points respectively. We did not impose the electron-phonon mediated spin mixing in any bands as these are found to be insignificant in similar structures \cite{Molina2015}. We reserve our discussions on the Eliashberg functions shown in Figs. (2 c-e) in conjunction to the exciton formations in subsequent paragraphs. The SFs of valance and conduction bands at $\mathsf{K}$ are shown in Figs. (3 a-b). Since SFs are proportional to the FWHM of the polaronic linewidths, a sharper SF would then mean a more stable electronic state, i.e., a finite yet longer lifetime. As the temperature increases, the SF broadens in energy, thereby interacting much with phonons and leading to a fast non-radiative recombination. The SF broadening of the electronic states at 0 K can now be calculated from Eq. (4) in which the peak corresponds to the zero-point energy. One can then see a generic trend of red (conduction) and blue (valance) shifting of the peaks with temperature. Such a trend leads to the well-known band-gap shrinkage and is shown in Fig. (3-c). A careful analysis on the polaronic corrections to the width is shown in Fig. (3-d). The width varies slowly at lower temperature but increases linearly with at higher temperature. This behaviour is due to the renormalization of the electronic energies by acoustic phonons and is also suggested by the Eliashberg functions shown in Fig. (2-c) where the contribution from optical phonons is very less for the said bands. The residual width is the zero point correction and is found less at top valance but an order more in bottom conduction band at $\mathsf{K}$. This is in accordance with the uncertainity principle as smaller width results in a longer lifetime. We find that using a full dynamic self-energy computation, the conduction and valence band shrinks by 2$\%$ and 0.1$\%$ leading to a correction of 31 meV below the bare gap of 1.26 eV. The zero temperature QP weights for conduction and valance bands are found to be 95$\%$ and 98$\%$ leading to the conclusion of good QP state. The SF peaks of later are much larger in magnitude than the former due to the larger widths. Increasing temperature to 1000 K reduces the QP weights to 73$\%$ and 91$\%$ respectively, questioning the validity of the QP assumptions for the conduction band. However even at this temperature, we did not find any appreciable emergence of secondary peaks as in case of trans-polyacetylene \cite{Cannuccia2016}.\\
\begin{figure*}[!ht]
  \includegraphics[width=1.5\columnwidth]{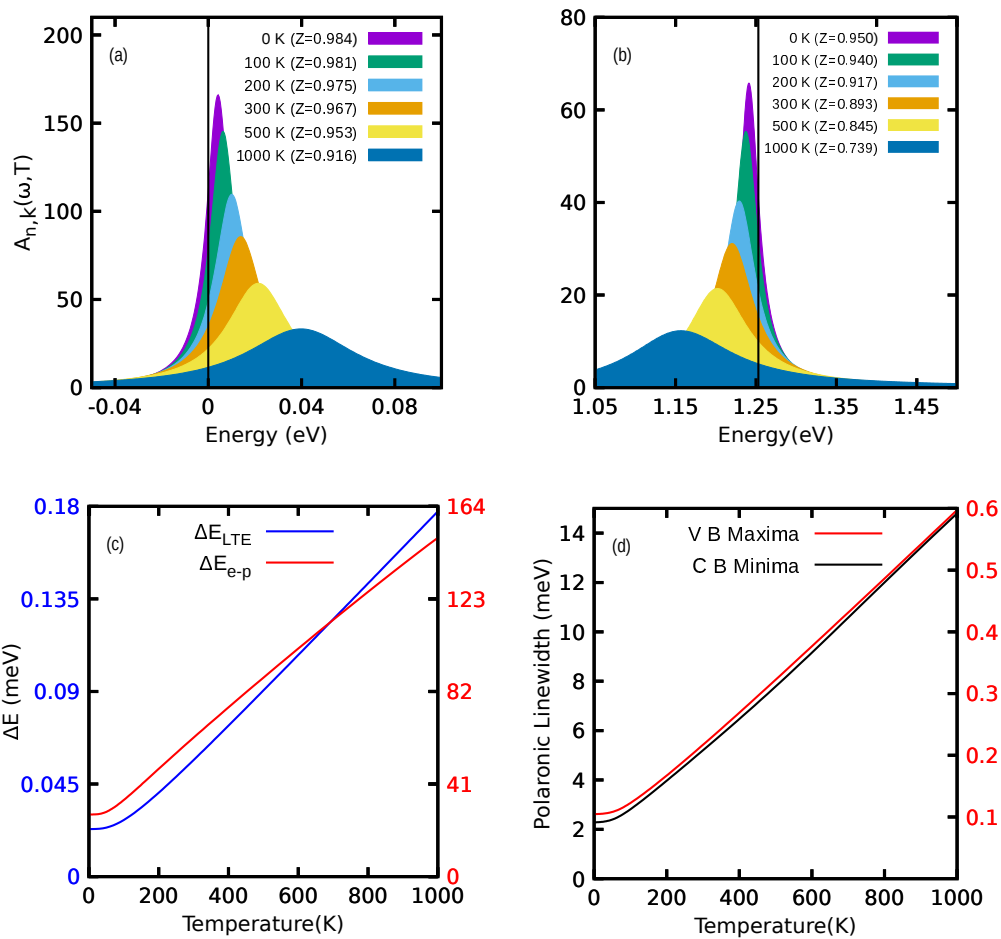}
  \caption{(a-b) Spectral functions at the top valance band and lowest conduction band at different temperatures respectively. The vertical line marks the bare energy values. The Lorentzian shape broadens as temperature rises with respective blue and red shifting of the valance and conduction band. (c) Zero point energy as function of temperature due to the electron-phonon renormalization (red curve) and lattice anharmonicity (blue curve) representing a band gap reduction with temperature. (d) Band resolved polaronic linewidth as function of temperature. Only the valance band maxima and conduction band minima linewidth at $\mathsf{K}$ point is shown.}
  \label{fgr:example}
\end{figure*}
To understand the band-gap reduction due to LTE (Eq. (6)), a QHA computation was further carried out. The variable volume QHA computations were performed on the same DFT charge-density results with the code developed by A. Dal Corso \cite{Giannozzi2009} under the same $\mathtt{PWscf}$ code sub-routines. The lattice anharmonicities are captured by altering only the in-plane lattice constants to two geometries centred about the relaxed one with a strain of $\pm$0.01 $\mathring{\mathrm{A}}$. Once the Helmholtz free energy is minimized, the corresponding zero-point energy was extracted by fitting into the second order Birch-Murnaghan equation of state \cite{Mounet2005}.         
\begin{figure}[!h]
  \includegraphics[width=1.0\columnwidth]{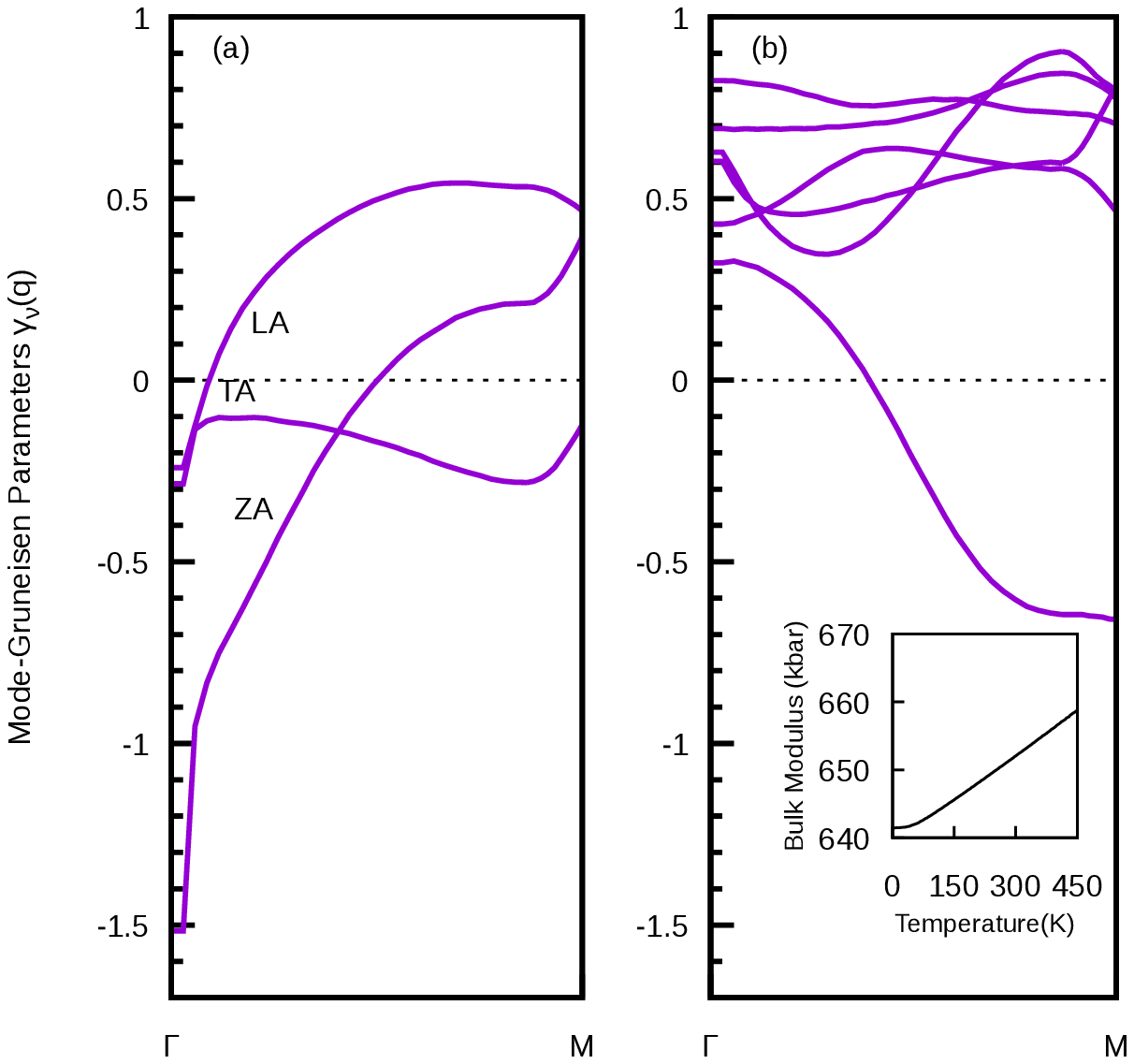}
  \caption{The Mode dependent \textnormal{Gr\"uneisen} parameter along the $\Gamma$-M BZ route. The lowest three acoustic modes are shown in (a) while the optical modes are shown in (b). A negative $\gamma_{\lambda}\left(\textbf{q}\right)$ is obtained in ZA mode at $\Gamma$ resulting in a lattice thermal contraction. The inset shows the variation of the bulk modulus with respect to the temperature.}
  \label{fgr:example}
\end{figure}
\begin{table}
  \caption{ZPR of ML TMDCs and other materials}
  \label{tbl:notes}
  \begin{tabular}{ll}
    \hline
    Structure                            	& ZPR (meV) \\
    \hline
    ML WSe$_{2}$   							& 31 (This work) \\
    Black Phosphorus \cite{Villegas2016} 	& 20 \\
    trans-polyacetylene \cite{Cannuccia2016} & 40 \\
    ML MoS$_{2}$ \cite{Molina2015} & 75 \\
    Si \cite{Monserrat2014} 					& 123 \\
    SiC \cite{Monserrat2014} 				& 223 \\
    polyethylene \cite{Cannuccia2016} 		& 280 \\
    Diamond \cite{Antonius2014} 				& 622 \\
    \hline
  \end{tabular}
\end{table}
To understand the anharmonicities that might contribute to the ZPR at a non-zero temperature, we additionally computed the \textnormal{Gr\"uneisen} parameter $\gamma_{\lambda}\left(\textbf{q}\right)=-\frac{\partial\ln\omega_{q\lambda}}{\partial\ln V}$ for each of the phonon mode along the $\Gamma$-M route of the BZ. The lowest three acoustic modes are shown in Fig. (4-a) whereas the rest optical modes are displayed in Fig. (4-b). Immediately by looking at the centre $\Gamma$ of the BZ, we find that the out-of plane ZA mode appears to be the dominant mode that dips to a much negative value resulting in a lattice thermal contraction at lower temperature. The optical modes which are not yet excited are clear signatures of positive $\gamma_{\lambda}\left(\textbf{q}\right)$. As the temperature rises, the dormant optical modes wake up and lattice expansion occurs. The lattice contraction in this case however is indeed small and was captured by \textnormal{\c{C}}ak{\i}r, et. $al$. \cite{Cakir2014}. Since there is no stacking layer involved in ML WSe$_2$, a positive strain then increases the ZA frequency of vibrations resulting in the well-known membrane-effect \cite{Landau2010}. This was earlier observed in graphene \cite{Mounet2005} and recently on silicene, germanene and blue phosphorene \cite{Ge2016}. Interestingly, we obtained a $\gamma_{A_{1g}}\left(\Gamma\right)$ = 0.69 which is close to the experimentally evaluated value of 0.55 \cite{Yang2017}. The computation of ZPR uses the bulk modulus $B_{T}$ and therefore, in the inset of Fig. (4-b) the variation of the same with temperature is shown. We find a monotonically increasing behaviour with a room temperature $B_{T}$ = 65.2 GPa, which is a little less than it's corresponding bulk value of 72 GPa \cite{Yang2017}. The influence of LTE on band gap reduction is then shown back in Fig. (3-c). The anharmonic contribution is small, as expected due to the larger atomic masses, however does not show any decreasing (anomalous) behaviour. We thus find that the gap shrinkage is mainly due to the electron-phonon interactions. Table 2 summarizes the ZPR obtained in this work with other ML TMDCs and bulk members. Comparing with the ML counterpart, we see that the ZPR decreases as the weight of the formula units become more which is consistent since heavier atoms vibrate less.
\begin{figure*}
\includegraphics[width=1.5\columnwidth]{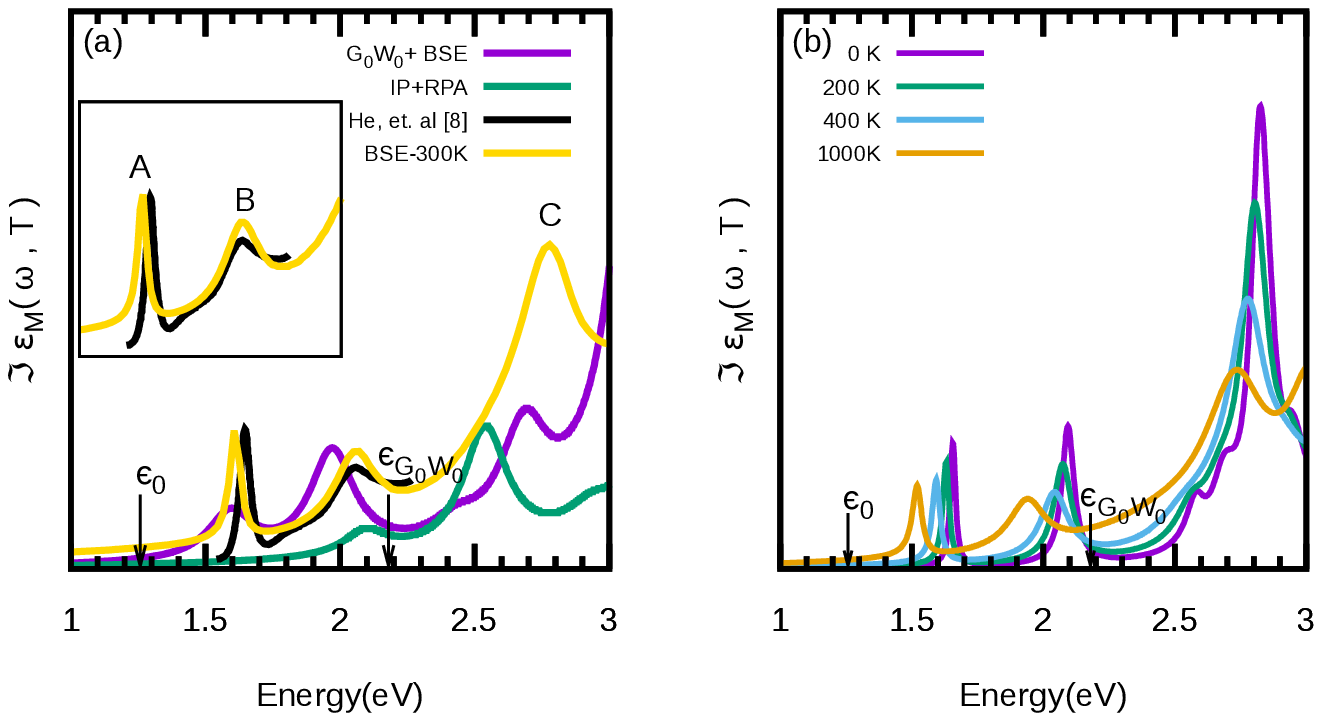}
  \caption{BSE is used to extract the temperature dependent absorption spectra as function of photon energy at (a) 300 K. The peaks A and B and C corresponds to the neutral excitons. Both the excitons A and B are within the bare energy and continuum, marked by arrows. Black curve is the experimental data \cite{He2014}, also shown as an enlarged manner in the inset. The frozen-atom approximated spectra is shown in purple color, while the spectra corresponding to the independent particle approximation under RPA and local fields is shown by green line. (b) The absorption spectra at different temperatures exhibiting a relative blue shifting of the peaks.}
  \label{fgr:example}
\end{figure*}
The absorption spectra at different temperatures has been computed using a coupled BSE and is shown in Fig. 5. Both the resonant and antiresonant electron-hole pairs were included in the BS kernel since the omission of the later, known as the Tamm-Dancoff approximation \cite{Myrta2009}, underestimates the collective density oscillations which consists also the electron-hole pairs. The absorption spectra at 300 K has been exhibited in Fig. (5-a) which contains the polaronic widths and quasi energies at each state $\left|n,\textbf{k}\right\rangle$ as well as the dynamic long range electron-electron gap correction. To include this G$_{0}$W$_{0}$ correction, we use a scissor operator of 0.928 eV and a linearly fitted conduction and valance band stretching of 1.134 eV and 1.069 eV respectively (see the Supplementary Information). The vertical arrows represents the bare energy gap 1.26 eV and G$_{0}$W$_{0}$ gap 2.19 eV. A static screening was then computed to build up the BS kernel. The polaronic and G$_{0}$W$_{0}$ corrections makes the BS matrix a non-Hermitian. Therefore a full-diagonalization method is necessary to solve for the excitonic strengths and lifetime which included top 6 valance and 5 lowest conduction bands for transitions. The value of the response block size is kept same as in the case of G$_{0}$W$_{0}$ calculation i.e., 5 Ry while the electric polarization vector direction was chosen to be normal to the plane of ML. The above configuration was found sufficient for optimizing the computational resources to a bare minimum, yet producing accurate results. The inset in Fig. (5-a) shows an enlarged view for comparison of our work with the reported experimental results \cite{He2014}. The A and B excitons are found to be located at 1.61 eV and 2.06 eV compared to the experimental value of 1.65 eV and 2.08 eV. The ground state exciton A is formed as a result of inter-band transitions from $\mathsf{V_{1}}$ valence energy band to the two lowest conduction bands $\mathsf{C_{1}}$ and $\mathsf{C_{2}}$ at $\mathsf{K}$ ($\mathsf{K}^{\prime}$) points in the BZ (Fig. (1-d)). Similarly, exciton B is build due to the electronic transition from $\mathsf{V_{2}}$ to $\mathsf{C_{1}}$. The difference between these two excitonic peak is exactly equal to the bare SoC valence bandwidth. Apart from these two peaks the C-exciton is located at 2.77 meV and is made from the deep-lying conduction and valance bands that take part in inter-band transitions resulting in a spectrally broad response including contributions near the $\Gamma$ point. Eliashberg function can now be used to understand the coupling of the phonon modes with the three excitons. We now return to Fig. (2 c-e) in which we computed the difference between the conduction and valance Eliashberg functions and is calculated for those bare states where the transitions has taken place. For example, $\mathsf{V}_{1}\rightarrow \mathsf{C}_{1}$, $\mathsf{C}_{2}$ means the transition has occurred between top valance band and two lowest conduction bands (at $\mathsf{K}$) (Fig. (1-d)), and is the case with A exciton. The difference is then $\Re\left[g^{2}F_{C_{1}}\left(\omega\right)-g^{2}F_{V_{1}}\left(\omega\right)\right]$. This quantity, being a negative for semiconductors without any anomalous band gap shrinkage with temperature, in this case is found to be dominated mainly by the acoustic branches around 103 cm$^{-1}$. Out of these three, the LA branch is found to be the most significant one resulting in an in-plane torsional mode at $\mathsf{K}$. The other acoustic branches contribute to only stretching of the lattice along the in-plane and out-of plane directions. A small peak around 255 cm$^{-1}$ is due to the optical branch consisting of both in-plane (W atoms) and out-of plane (Se atoms) torsional modes, but effective only by about 25$\%$ compared to the former. This is in stark difference found in MoS$_{2}$ ML A-exciton build-up where the main contribution comes from the optical branch around 400 cm$^{-1}$ \cite{Molina2015, Saigal2015}. The B exciton couples in a similar manner with almost one-third contribution from optical branches. A small addition can be seen coming from the mid-frequency branches and is shown by a small peak around 194 cm$^{-1}$. Interestingly, C-exciton couples with both acoustic (around 116 cm$^{-1}$) as well as lower and higher optical branches (around 210 cm$^{-1}$ and 244 cm$^{-1}$) and is 70$\%$ more effective compared to the in-plane contribution. The effect of varying temperature on absorption spectra is shown in Fig. (5-b) where a more red shifting of the absorption peaks is obtained. Analysing carefully the intensity and broadening of excitons, we find that the B peaks more than A in the frozen-atom approximation (only G$_{0}$W$_{0}$ correction) and at 0 K. However, as the temperature increases, both the peaks start becoming smaller in magnitude. At all temperatures, the A peak is found to be slightly narrower than the B peak, but the spectra magnitude reduces more for B exciton. Furthermore, both A and B excitons remains unchanged in terms of intensity but there is a considerable broadening specially for B exciton. These results are in excellent agreement with the experimental results where the PL intensity of the peaks contributing to the lower bound excitons A and B decreases with increasing temperature \cite{Schmidt2016}. Here we could not capture the higher lying exciton as demonstrated by Schmidt et. $al.$ \cite{Schmidt2016}, which follows an opposite trend with temperature after 110 K. This could be due to its existence as a result of inter-valley transition which stands as a limitation in this current BSE execution. In their experiment they studied the nature of all excitons and trions and found that only this higher lying neutral exciton shows such an anomalous behaviour with temperature. Additionally, they demonstrated that due to the dominance of this particular neutral exciton's intensity at higher temperatures, the integrated PL intensity (i.e., sum of contribution of all excitons and trions formed when the sample is exposed to light) shows an increasing trend with temperature as also shown elsewhere \cite{Zhang2015}. All the other effects as shown in this work such as the red shift of the peaks in the absorption spectra with increasing temperature and width of the peaks of the individual neutral excitons follows the same trend as demonstrated elsewhere \cite{Schmidt2016, Zhang2015}.\\
The C-exciton with it's peak located at 2.77 eV is is good agreement to the reported experimental value spread about 2.5 eV \cite{Ahna2017}. A sight discrepancy in our result is might be due to the use of scissor operator in BSE calculation that stretches the conduction and valance bands and therefore underestimates the energy coordinate of the higher lying excitons. Whatsoever, we find that C behaves differently with increasing temperature having its intensity reduced at a higher pace and increased broadening considerably. In addition to these, the absorption spectra in the absence of BS kernel (i.e., independent particle approximation) but including the static screening within the RPA has been plotted to compare the consequence on the peak positions. The oscillator strengths of the three excitons over a range of temperature is shown in Fig. (6-a). Excitons A and B are found to remain bright at all temperatures, whereas there is only a slight variation in the strength in case of C-exciton in the range 250-1000 K. This can be understood from the first- and second-term in Eq. (10). In case when $\left|\left.\varphi\left(T\right)\right\rangle \right.\sim\left|\left.\varphi_{FA}\right\rangle \right.$, the second term dominates and as a consequence there is an individual interaction that occurs between the phonons with electrons and holes. This is therefore the incoherent contribution and is the case with ML WSe$_{2}$.
\begin{figure*}[!ht]
  \includegraphics[width=1.5\columnwidth]{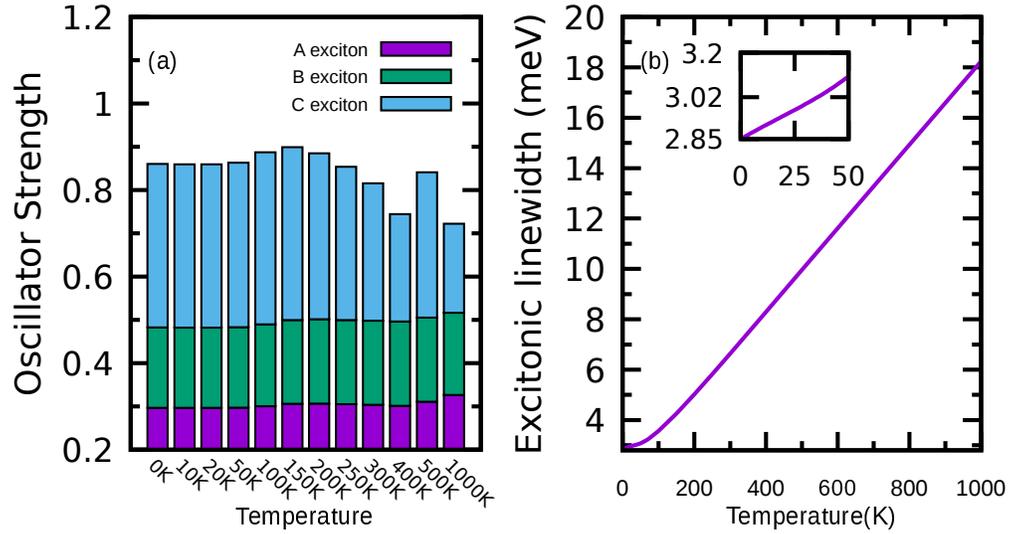}
  \caption{(a) Exciton oscillator strengths at different temperatures. All the excitons are found to be bright at all temperatures. (b) A-excitonic non-radiative linewidths as function of temperature. The inset enlarges a low temperature variation.}
  \label{fgr:example}
\end{figure*}
A bright to dark transition or vice versa would resulted in a non-zero but large value of the first-term and hence is the coherent contribution. Such transitions occurs due to transfer of oscillator energies when a bundle of excitonic states get close, and is found to dominate in h-BN \cite{Marini2008}. As can be seen from the said figure that there is no excitonic states close in energy either for A, B or C excitons hence there is no bright to dark or vice-versa transition for any of the excitons. The A-exitonic non-radiative linewidth as function of temperature is shown in Fig. (6-b). 
\begin{figure*}
  \centering
  \includegraphics[width=1.5\columnwidth]{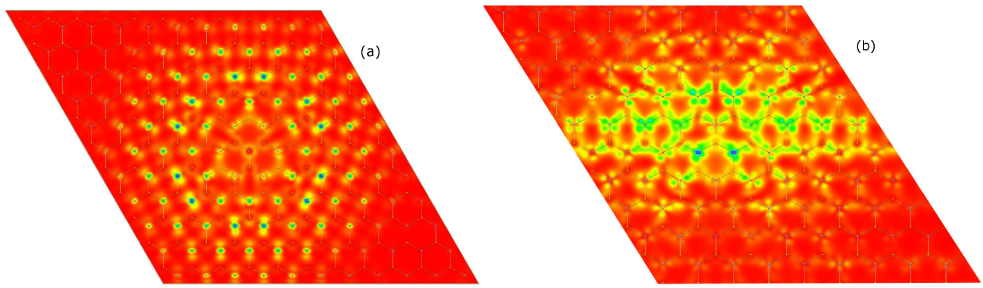}
  \caption{Excitonic wavefunction plot at 300 K over the $xy$ plane of the ML WSe$_{2}$. The hole is fixed in space and is placed on the top of the W atom at a distance of 1.05 $\mathring{\mathrm{A}}$. The wavefunction spreads to about (a) 31.60 $\mathring{\mathrm{A}}$ over the entire plane in case of A exciton. For (b) C- exciton, the spreading is more bright with a large oscillator strength and spreads only upto 26.56 $\mathring{\mathrm{A}}$.}
  \label{fgr:example}
\end{figure*}
The residual linewidth at 0 K is found to be 2.85 meV. The linear section can be explained by fitting to the phonon induced variation $\varUpsilon\left(T\right)=\varUpsilon_{0}\left(0\right)+\varUpsilon^{\prime}T$, where $\varUpsilon^{\prime}$ denotes the linear increase due to the acoustic phonons. An exponential variation would resulted in an exciton-optical phonon interaction, which is recently found to be present in ML MoS$_{2}$ \cite{Cadiz2017}. When extrapolated to 0 K, we find $\varUpsilon_{0}\left(0\right)$=2.25 meV. This is close to the value 1.6$\pm$0.3 meV in case of a chemical vapour deposition (CVD) grown ML WSe$_{2}$ where the exciton-phonon interaction is mainly found to be stemming from the acoustic modes \cite{Moody2015}. Furthermore, we obtained the exciton-acoustic phonon coupling strength $\varUpsilon^{\prime}$ = 15.6 $\mu$eVK$^{-1}$. This is almost four times smaller than the CVD samples. The reason for these limitations in our results is might be due to the ruling out of any defects in the lattice, exciton-exciton/trion interactions (i.e., incorporation of higher order Feynmann diagrams and formulation of three-body time dependent DFT Hamiltonian) as such. Interestingly, in Figs. 7(a) and 7(b), we have plotted the A and C-excitonic wave functions at 300 K over the in-plane coordinates respectively. The wave functions in both the case was calculated by fixing the hole position above the W atom at a distance of 1.05 $\mathring{\mathrm{A}}$. We took this distance since we found the electron density to be higher near the Se atoms and is also shown in the electron density plot (see Supplementary Information). The excitonic wave-functions were then unfolded on the real space lattice by using $\mathtt{Xcrysden}$ code \cite{Anton2003}. The spreading of the wave function of A exciton is over a diameter of 31.60 $\mathring{\mathrm{A}}$ and therefore envelopes many unit-cells. They are thus the Wannier-excitons. Moreover, the  wave function appears to be spherically symmetric suggesting a 1s ground state, similar to that of the ground state wave function of hydrogen atom, is properly exhibited. The wave function of B exciton is found to be of similar nature and hence not shown here. A rather more localised with high binding energy of C-exciton is shown in Fig. (7-b) which spreads over a diameter of 26.56 $\mathring{\mathrm{A}}$. These spreading of the excitons are smaller than that of the ML MoS$_{2}$ obtained by $ab$-$initio$ results \cite{Molina2015}, confirming a higher excitonic binding energies at room temperature in ML WSe$_{2}$. We understand that inclusion of inter-valley transitions explaining the intriguing spin as well as momentum forbidden low lying dark states together with the thermal expansion effects would have been much accurate in describing the quenching of the PL spectra at low temperatures. Such a computation would require a solution of the time dependent BSE and is beyond the scope of this work. However, in this work we have presented an extensive quantitative analyses in order mainly understand the essential phonon couplings between the A, B, C excitons and the phonons. The role of lattice anharmonicities towards the zero pint gaps are also investigated.
\section{Conclusions}
The lack of inversion symmetry and reduced screening in ML TMDCs plays a central role in building up a strong spin-orbit interaction hosting bound excitons with large binding energies. In this work, we demonstrate a contemporary and combined electron-electron, electron-hole and electron-phonon study to reveal a thorough underlying mechanism of exciton-phonon couplings and energy renormalization in ML WSe$_{2}$. We use a complete $ab$-$initio$ formalism (starting from the bare energies) to explain the origin of neutral exciton-phonon couplings and the respective excitonic linewidths. The absorption and excitonic energies were obtained by solving the coupled electron-hole BSE that included the polaronic energies extracted from the DFPT calculations and are found to be in excellent agreement to the reported experimental data over a wide range of temperature. Within this approach we were able to capture features, such as the oscillator strength of the excitonic peaks and their broadening, directly related to the temperature and to the non-radiative exciton relaxation time. Eliashberg functions were computed from these results and were analysed against the phonon spectrum to understand the mode coupling with the electronic transitions between different energy bands about the optical gap. We found that the A and B excitons mainly couples with the LA phonons. C-exciton couples both with LA as well as the optical modes near 225 cm$^{-1}$. Apparently, we also found that the electron-phonon interaction strongly renormalizes the bare electronic energies. To quantify the contributions from lattice anharmonicities, we executed a variable volume quasi-harmonic analysis of the mode dependent \textnormal{Gr\"uneisen} parameter along the BZ route. This resulted that the out-of-plane mode contributes mostly to the lattice thermal contraction at lower temperature. Albeit of this, we find that the effect of the lattice anharmonicities imparts much less energy corrections to the optical band gap compared to the electron-phonon interactions. We believe that this work for a reference would be very useful to study the behaviour of ML WSe$_{2}$ based excitonic solar cells at finite temperatures where the exciton physics governs the photo-conversion.

\begin{acknowledgments}
SB acknowledges the financial supports from the DST, India with the grant number YSS/2015/000985 under Fast-Track Young Scientist Programme and grant number IFA 12-ENG-39 under INSPIRE Faculty Award Programme. HM and AB acknowledges MHRD and DST, Govt. of India, for financial education support respectively.
\end{acknowledgments}
\nocite{*}

\bibliography{apssamp}

\end{document}



\title{Supplementary Materials for:\\Exciton-Phonon Coupling and Band-Gap Renormalization in Monolayer WSe$_2$}
\author{Himani Mishra${}^\ddagger$}
\author{Anindya Bose${}^\ddagger$}%
\author{Amit Dhar$^{\dagger}$}
 \author{Sitangshu Bhattacharya${}^\ddagger$}
 \email{Corresponding Author's Email: sitangshu@iiita.ac.in}
\affiliation{${}^\ddagger$Nanoscale Electro-Thermal Laboratory, Department of Electronics and Communication Engineering, \\ $\&$ \\
${}^\dagger$Department of Information Technology, \\
Indian Institute of Information Technology-Allahabad, Uttar Pradesh 211015, India}
             \pacs{71.35.-y, 71.35.Cc, 63.20.Dj, 63.20.Ls., 31.15.Md, 11.10.St}
\maketitle
\onecolumngrid
\newpage \section{Kinetic energy cut-off convergence}
\begin{figure*}
  \includegraphics[width=0.75\columnwidth]{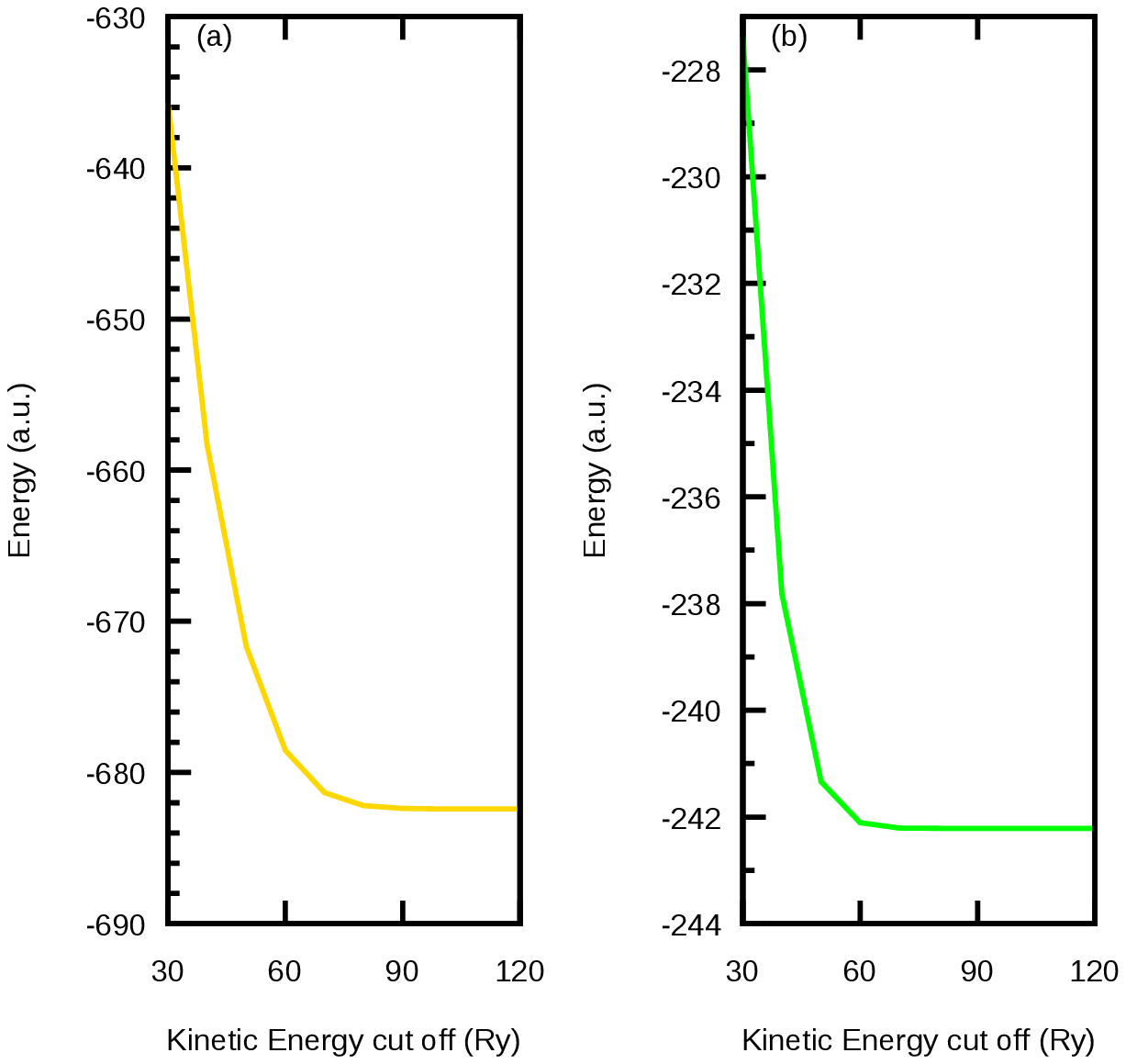}
  \caption{Total energy minimization as function of kinetic energy cut-off for a norm-conserving, core corrected pseudopotential of (a) W and (b) Se. We see that energy settles from 90 Ry in both cases. We therefore choose this value in all our computation.}
  \label{fgr:example}
\end{figure*}

\newpage 
\section{k-point sampling convergence in DFT}
\begin{figure*}
  \includegraphics[width=0.95\columnwidth]{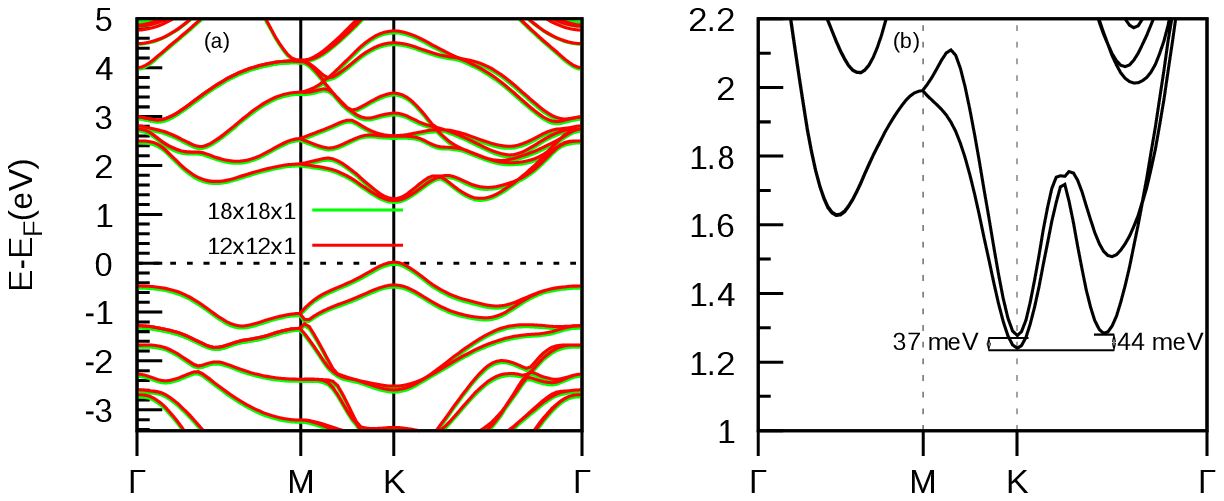}
  \caption{Effect of (a) 12$\times$12$\times$1 and 18$\times$18$\times$1 on the band gap convergence along the symmetry lines of the BZ route in ML WSe$_{2}$. (b) Lowest conduction band edges along the same BZ route done on signifying a splitting of conduction bands. Both the sampling resulted a direct band gap.}
  \label{fgr:example}
\end{figure*}

\newpage \section{G$_{0}$W$_{0}$ QP energy stretching as function of bare energy}
\begin{figure*}
  \includegraphics[width=0.75\columnwidth]{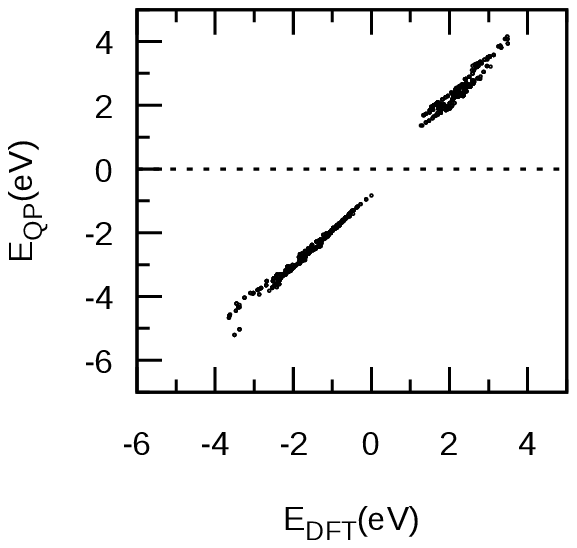}
  \caption{G$_{0}$W$_{0}$ QP energies as function of DFT energies. The top of the DFT valance band energy is located at 0-level. Now in order to account for the deviation between DFT band structure and G$_{0}$W$_{0}$ band structure, we apply a data regression technique and calculate the spreading of the bands both in the valence band and conduction band regions which comes around 1.134 eV for conduction bands and 1.069 eV for valence bands. This is necessary along with the incorporation of the scissor operator so as to consider any shift in the bands other than VBM and CBM.}
  \label{fgr:example}
\end{figure*}

\newpage \section{Electron density contour around Tungsten and Selenium atom}
\begin{figure*}
  \includegraphics[width=0.85\columnwidth]{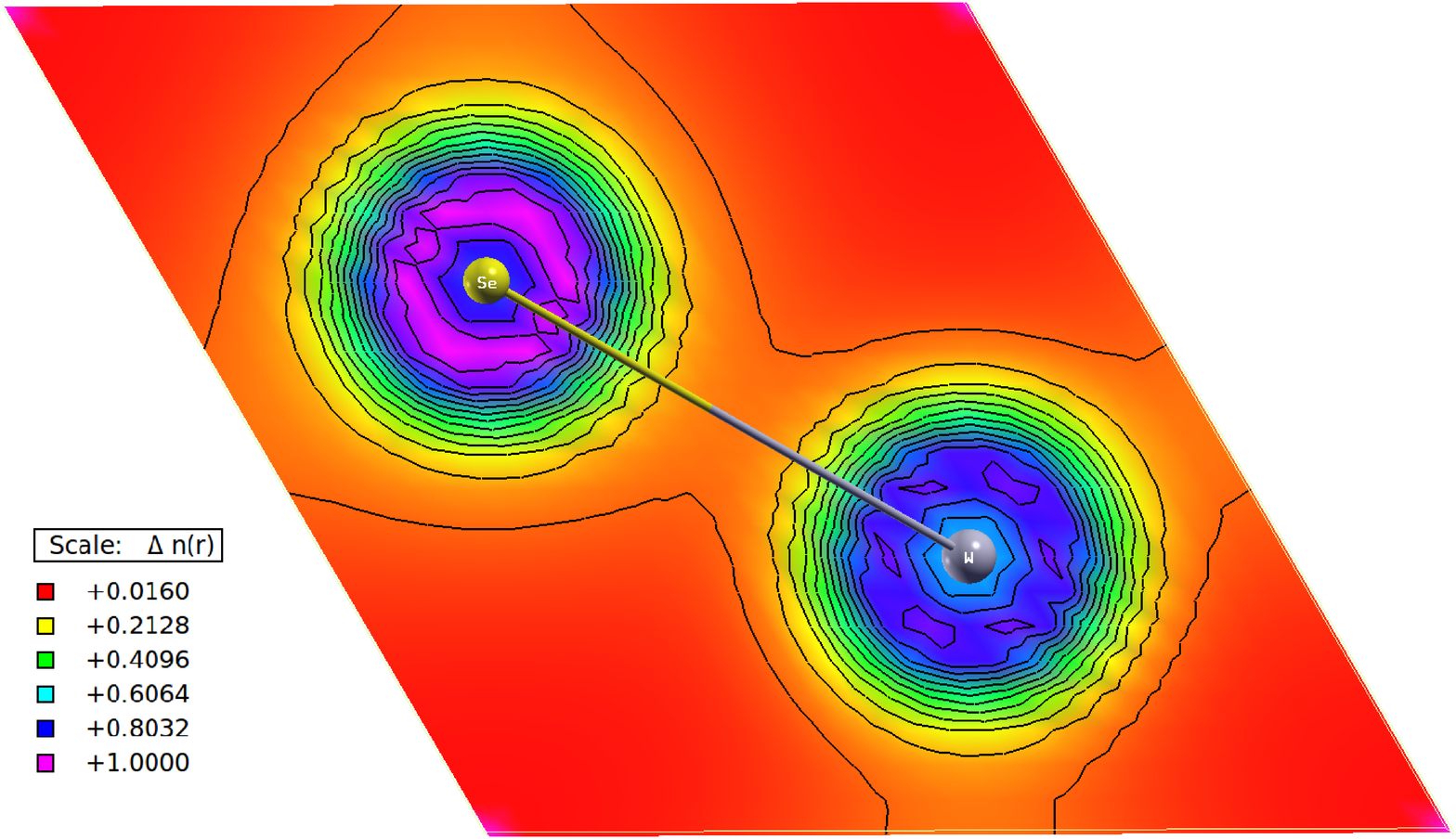}
  \caption{Electron density contour over W and Se atom. This shows the electron wave function i.e., the measure of the probability of an electron being present at a specific location. It can be clearly seen from the plot that this probability is highest around the Se atom. Due to this reason, we have placed the hole above the W atom in the analysis of exciton wave function, as it has high probability of hole existence.}
  \label{fgr:example}
\end{figure*}

%
\bibliography{}